# A Survey of the Individual-Based Model applied in Biomedical and Epidemiology Research

Erivelton G Nepomuceno[1*]
Denise F Resende[1]
Márcio J Lacerda[1]

[1]Department of Electrical Engineering, Federal University of São João del-Rei, Brazil

## Abstract

Individual-based model (IBM) has been used to simulate and to design control strategies for dynamic systems that are subject to stochasticity and heterogeneity, such as infectious diseases. In the IBM, an individual is represented by a set of specific characteristics that may change dynamically over time. This feature allows a more realistic analysis of the spread of an epidemic. This paper presents a literature survey of IBM applied to biomedical and epidemiology research. The main goal is to present existing techniques, advantages and future perspectives in the development of the model. We evaluated 89 articles, which mostly analyze interventions aimed at endemic infections. In addition to the review, an overview of IBM is presented as an alternative to complement or replace compartmental models, such as the SIR (Susceptible-Infected-Recovered) model. Numerical simulations also illustrate the capabilities of IBM, as well as some limitations regarding the effects of discretization. We show that similar side-effects of discretization scheme for compartmental models may also occur in IBM, which requires careful attention.

## Introduction

For centuries, humanity has been marked by adversity in the pursuit of survival. Wars, famine, the climate beyond predators are some of main challenges for humanity's progress and survival. However, no other factor brings so much fear to society as epidemics [1]. The black death of the fourteenth century and the influenza pandemic that occurred during First World War were responsible for at least 100 million deaths [2,3]. Nowadays, acquired immune deficiency syndrome (AIDS) has been universally recognized by its significant impact on mortality rates, especially in developing countries, threatening global health. In 2016, an estimated 36.7 million people were living with AIDS [4], were 19.4 million are in East and Southern Africa. This is an example of the disastrous effect of infectious diseases. Although the application of public health polices has controlled some main epidemic threats, particularly in developing countries, policies to limit their occurrence have been insufficient [5,6]. The course of the epidemic is determined by the social aspects and the initiatives implemented to contain it.

Therefore, understanding the epidemiological process and the pathogenesis of diseases has became increasingly relevant [7]. The study of the dynamics of infectious diseases, as well as methods that may support the development of strategies for prevention and control of diseases through mathematical models, have been widely developed. Mathematical models provide a powerful toolkit in the process to control and prevent the emergence, expansion or resurgence of pathogens, since the availability of options requires continuous evaluation using different methods. Also known as mathematical epidemiology, this tool is one of the most important scientific activities to help in the eradication of diseases [8].

Among many mathematical modeling strategies, one of the most studied is the compartmental technique, such as the SIR model (Susceptible-Infected-Recovered) [9]. The SIR model has been applied in many areas [10]. Using a time series SIR model, the authors in [11] have found that the initial phase of the epidemic is stochastic. Stone et al. [12] and Aron and Schwartz [13] obtained period-doubling bifurcation and chaos in epidemic models. Nowak et al. [14] indicated that mathematical models are used for understanding phenomena like AIDS pathogenesis. There are many papers that apply control theory in several compartmental models. In general, in the theory of epidemic dynamical models there are continuous-time models described by differential equations, and the discrete-time models described by difference equations. However, differential (difference) equations are not adequate to model a system in which the individuals present important differences [15]. This behavior is due to the fragility of compartmental models assuming that the distribution of individuals is spatially and temporally homogeneous. Therefore, one of the most prominent strategies is to deal with an individual as a single entity, that is, a heterogeneous mixture of individuals. This is a feature of the Individual-Based Model (IBM) [16].

Individual-Based model (IBM) has been used to analyze epidemiological mathematical models in a more realistically way. The IBM deals with many entities, spatial scales, heterogeneities, and stochastic events, it is necessarily more complex than classical models [17]. Thus, IBM's purpose is to model infectious diseases in populations in which individuals can be divided into epidemiological states. This framework is intended as an alternative to replace or complement compartment models, such as SIR [5]. More recently, a considerable number of works have been developed to model and control disease around the world using IBM as a tool, strategies such as vaccination programs or isolation programs, such as School closure, are applied to mitigate the effects of a pandemic. Due to its excellent results, vaccination is still considered one of the most effective strategies to prevent infectious diseases and their consequences [18]. Most often, the IBM simulates experiments in a computational environment, considering individual characteristics and the environment in which they are, i.e. interactions between individuals [19]. The implementation of such models is already facilitated by software platforms specifically designed to implement models based on individuals [20,21]. Future perspectives for the IBM can be assembled from a tried and tested construction block toolkit representing, for example, energy budgeting, habitat selection, and feature composition [22]. In this paper, a review section investigates the frequency and methods of such applications in IBM. The focus of this survey is on IBM applied to infectious diseases in humans and their perspectives [23,24]. In this systematic review, we summarized and discussed IBM applications and terminology in different epidemiological disciplines, published between 1997 and 2017. The characteristics of the model were analyzed and discussed, such as the implementation of social mixtures, demographic evolution over time, modeling platforms for IBM's, and such as strategies to contain an epidemic. In addition to the literature review, three simulation experiments will be presented. First, an example of IBM is described. Second, IBM is adjusted to have an average behavior corresponding to a SIR model. In the third experiment, the limitations of the models related to the effects of discretization and the generation of chaos are presented.

## Material and Methods

### Literature Review

A systematic review of studies using IBM applied to the transmission of infectious diseases in humans has been conducted. The strategy adopted for review research is to use IBM as the comprehensive term for individual-level models, also observed as agent-based model, cellular automata (CA), and so on. The definition based on the literature for Individual-Based Model is: "A computational tool capable of simulating individuals that interact with each other, where individuals or agents have unique attributes that change throughout the life cycle" [21,22].

**Search:** This is a literature review with search frequency from 1997 to the present. The search engines were SCOPUS, IEEE and WEB OF SCIENCE. Based on the listed definitions and exploratory searches, the following search query was used: "("individual based model") AND (biomedic* OR "infectious disease" OR epidem*)". With date and descriptor restrictions, initially there were 226 articles in the Web of Science database, where the Topic search restriction was used meaning search by title, abstracts, and keywords in all databases. For the Scopus database, 233 studies were found by searching by title, abstracts and keywords in all databases. For the IEEE database, 673 studies were found in all databases. Original research papers were included using an IBM focused on the transmission of infectious diseases in humans. That is, reviews and studies related to animal research, ecology and molecular biology were excluded. All abstracts of these papers were read and evaluated for inclusion or exclusion in the present review. After the detailed reading of the abstracts, application of the inclusion and exclusion criteria, there were 31 papers in the Web of Science, 55 in Scopus and 20 in the IEEE. From a total of 1132 studies, 106 papers were selected. After checking for some duplication, the final sample of the present review is composed of 88 articles. The search process is shown in Figure 1.

To the best of authors' knowledge, we could not find a survey of IBM with focus on biomedical and epidemiological research. This is one of the main objectives of this paper. Additionally, as it has been already mentioned, we have provided some general mathematical description of the IBM as well some concerns regarding the effects of discretization schemes, which are more usually investigated for differential equations, but they may also bring some inconsistent results for the IBM.

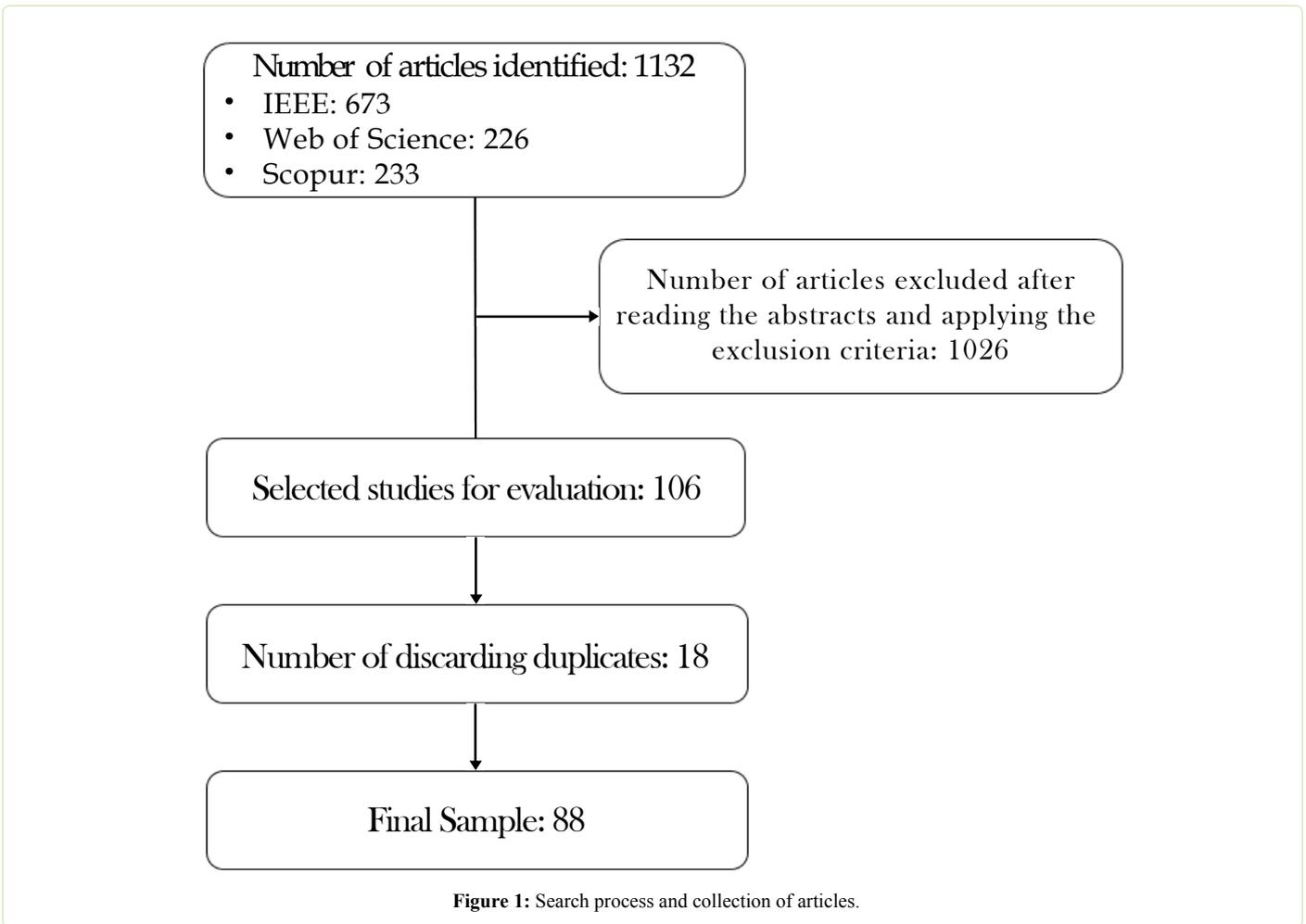

**Figure 1:** Search process and collection of articles.

**Model classification:** The articles were classified in three broad areas of approach:

**Topic:** Type of pathogenesis.

- **Disease:** The disease addressed in Article;
- **Type:** Forms of contagion;

**Modeling purpose:** The proposal presented.

- **Methods:** Description of new approaches to IBM research, introducing modeling concepts, performance enhancements, simulation techniques, or mathematical techniques;
- **Dynamics:** Using methods to understand the dynamics of the transmission and elaborate on the best assumptions of the model a viable solution;
- **Interventions:** Evaluation intervention measures to inform policy makers and organizations, based on knowledge of the dynamics of transmission and economic analysis;

**Control strategy:** Intervention strategy used.

- Vaccination;
- Antivirals, drugs;
- Screening;
- Non-pharmaceutical intervention (NPI): social distancing, closure of schools, improvement of living standards and urban mobility.

**Full-text trace:** To extract features and applications from the model, a full-text exam was done for the selected articles as a function of the human application on infectious diseases. Some studies have been found in this context on Influenza, Hepatitis A, Hepatitis B, Hepatitis C, HIV, HPV, Measles, Polio, Ebola, Smallpox, Chickenpox, SARS, Respiratory Syncytial Virus, Tuberculosis, Bubonic Plague, Pertussis, Lepra, Gonorrhea, Chagas Disease, Schistosoma, Dengue, Chikungunya, Malaria and Varicella Zoster. For each full-text article, we list the topic, the purpose of modeling and control strategy that have been used, employing IBM as a tool.

**Bibliometric indicators:** Table 1 presents schematically the results found giving an overview of the different themes, modeling purposes and study characteristics. The result of the review consists of 86 articles and the years of publication ranged from 1997 to 2018. The Search Process was explained in Section 2.1.1. during the selection of the articles it was possible to observe a change in the applications used in IBM, which explains the selection of more recent works, this change occurred from studies focused on methodology papers for applications and purposes related to intervention. Methodological papers, not applied to a specified close-contact infection, mostly describe the conceptual usage of an IBM to simulate heterogeneous disease dynamics. Among the selected papers, 2015 has the largest number of publications (17), one can consider this fact in front of the Ebola in 2015 [25-30]. From 1997 to 2003, no publication was found from the delineated search process.

**Table 1:** Characteristics of IBM studies applied to humans in the transmission of infectious diseases published from 1997 to 2018. Intervention studies may also describe dynamics and methods, as some papers may describe more than one strategy. HIV: human immunodeficiency virus, HPV: human papillomavirus, STI: sexually transmitted infection, SARS: Severe acute respiratory syndrome.

| Topic | | Count | Modeling Purpose | | | Control Strategy | | | |
|---|---|---|---|---|---|---|---|---|---|
| Disease | Type | | Methods | Dynamics | Interventions | Vaccine | NPI | Drugs | Screening |
| Unspecified Close-contact | General | 11 | 9 | 6 | 2 | 1 | 10 | - | - |
| Unspecified STI | General | 1 | - | 1 | - | - | 1 | - | - |
| Chickenpox | Viral | 1 | - | - | 1 | - | 1 | - | - |
| Ebola | Viral | 6 | 4 | 4 | 2 | 1 | 6 | 1 | - |
| Hepatitis A | Viral | 1 | - | - | 1 | 1 | 1 | - | - |
| Hepatitis B | Viral | 1 | 1 | - | - | - | 1 | - | - |
| Hepatitis C | Viral | 6 | 3 | - | 4 | 1 | 4 | 1 | 1 |
| HIV | Viral | 14 | 5 | 4 | 7 | - | 13 | 1 | 3 |
| HIV+Hepatitis C | Viral | 1 | - | 1 | - | - | - | 1 | - |
| HPV | Viral | 3 | - | 2 | 1 | 3 | - | - | 1 |
| Influenza | Viral | 13 | 8 | 3 | 6 | 2 | 10 | 2 | - |
| Measles | Viral | 1 | 1 | 1 | - | 1 | - | - | - |
| Polio | Viral | 2 | 2 | - | - | 2 | - | - | - |
| Respiratory syncytial virus | Viral | 1 | - | - | 1 | 1 | - | - | - |
| SARS | Viral | 1 | 1 | - | - | - | 1 | - | - |
| Smallpox | Viral | 2 | - | - | 2 | 2 | 2 | - | - |
| Varicella Zoster | Viral | 1 | 1 | - | - | 1 | - | - | - |
| Bubonic plague | Bacterial | 1 | - | 1 | - | - | 1 | - | - |
| Gonorrhea | Bacterial | 1 | 1 | - | - | - | 1 | - | - |
| Lepra | Bacterial | 3 | - | 1 | 3 | - | 1 | 2 | - |
| Pertussis | Bacterial | 1 | 1 | - | 1 | 1 | - | - | - |
| Tuberculose | Bacterial | 6 | 2 | 3 | 1 | - | 5 | - | - |
| Chagas disease | Parasitic | 1 | - | 1 | - | - | 1 | - | - |
| Schistosoma | Parasitic | 1 | 1 | 1 | - | - | 1 | - | - |
| Chikungunya | Vector-borne | 1 | - | 1 | - | - | 1 | - | - |
| Dengue | Vector-borne | 5 | 3 | 5 | - | 1 | 4 | - | - |
| Malaria | Vector-borne | 2 | - | 1 | 1 | - | 2 | - | - |
| TOTAL | | 88 | 43 | 36 | 33 | 18 | 67 | 8 | 5 |

Most papers in our selection are on HIV - human immunodeficiency virus (15.90%), closely followed by influenza (14.77%). Articles about influenza, according to the selection, use vaccination, isolation or school closure programs to control seasonal or pandemic influenza outbreaks [31-43]. On the other hand, the majority of the works of general close-contact diseases were written with the intention of describing the transmission dynamics or the methodology of mathematical models [44-53]. As previously mentioned, the applications of the IBM have migrated to a certain extent, from the use of methodological works for applications aimed at the intervention of infectious diseases. This is observed in the number of works related to human immunodeficiency (HIV) or sexual transmission [54-69], tuberculosis [70-75], dengue [76-80], Hepatitis C [81-86] and Ebola [25-30] in this review.

For vector-borne disease models, Dengue is the most studied subject. Generally, the studies try to understand the pathogenicity, transmission dynamics and even to create computational tools for the prevention and control of Dengue. Some papers have used computer platforms using mathematical software such as MATLAB [40,65]. In addition, software's such as NetLogo [52,74], AnyLogic [82], DengueME [78,79], HexSim [87] and NOVA [29] were used as explicit modeling platforms (Table 1).

Works published with IBM focused on infectious diseases are in constant ascendancy. In this review, most of the articles included were on topics focused on HIV, influenza or unspecified Close-contact, but this does not restrict studies on IBM. Studies of diseases transmitted by vectors, parasites [88,89] or bacteria [90-95] are also being approached and with a perspective of increasing work for the next years. Vector-borne diseases such as dengue [76-80], malaria [96,97] and chikungunya [98] are being improved, given the increasing geographic expansion of their vectors, generating future research perspectives.

Many papers use socioeconomic and biomedical databases to make predictions with mathematical models, along with interventions such as isolation and vaccination, to predict pandemic outbreaks, taking health care into account, and informing authorities of efficiency and efforts to end the pandemic. Few studies have been found on Chickenpox [99], Varicella [100], Hepatitis A [101], Hepatitis B [102], HPV [103-105], Measles [106], Polio [107,108], Respiratory Virus [109], SARS [110] and Smallpox [111,112], which do not mean that these pathogens are not being studied by other researchers because their socioeconomic aspects are extreme importance. Future research, using IBM, should address these and other topics more frequently, bringing some aspects that can influence the spread of diseases such as social awareness, climate change and global mobility.

**Mathematical models**

**Compartment models:** The SIR model (Susceptible-Infected-Recovered) is a compartmental model formulated in terms of differential equations, to study the consequences of a contagious disease that spreads rapidly in a population. The three compartments considered are:

- *Susceptible*: individuals who are not infected but may come to be.

- *Infected*: individuals who are with the disease and are able to pass it on to other individuals.

- *Recovered*: individuals who are recovered and immune to contagion. Immunity in this case consists of an individual with permanent immunity acquired by the infection.

There are other derivations of the SIR model presented to describe infectious diseases, following the same rationale. In order to obtain the set of equations that represent the SIR model, some considerations are made:

➢ Individuals are homogeneously distributed within a population.

➢ The population is considered constant, $d = \mu$.

➢ The number of individuals in the infected compartment increases at a rate that is proportional to the number of individuals in the infecting class and to the number of individuals in the susceptible compartment, mathematically represented by the portion $\beta(S)I(t)$.

➢ The rate at which infected individuals are transferred into the class of the recovered is proportional to the number of infected. This fact is modeled by $\gamma I(t)$. Thus, at an instant of time $t$ a population is characterized by:

$$N = S(t) + I(t) + R(t) \quad (1)$$

Where, $N$ is the total number of individuals in a population at instant $t$.

Therefore, the SIR model can be written as the set of differential equations:

$$\frac{dS}{dt} = \mu N - \mu S - \frac{IS}{N}, \quad S(0) = S_0 \geq 0,$$
$$\frac{dI}{dt} = \frac{IS}{N} - I - \mu I, \quad I(0) = I_0 \geq 0, \quad (2)$$
$$\frac{dR}{dt} = I - \mu R, \quad R(0) = R_0 \geq 0,$$

It is often set $s(t) = S(t) = N$, $i(t) = I(t) = N$ and $r(t) = R(t) = N$, as being, respectively, the proportion of susceptible, infected and recovered in a population of constant size $N$. $\beta$ is the infection rate, $\mu$ is the death rate and $\gamma$ is the recovered rate. Dividing the equations into (2) by $N$:

$$\frac{ds}{dt} = \mu - \mu s - \beta is, \quad s_0 \geq 0$$
$$\frac{di}{dt} = \beta is - \gamma i - \mu i, \quad i_0 \geq 0. \quad (3)$$

In this way, the population size is normalized, that is, $N = 1$, and the class of recovered can be determined by: $r(t) = 1 - s(t) - i(t)$

The fixed points of the Equations in (3) are obtained from:

$$\frac{ds}{dt} = f(s,i) = 0$$
$$\frac{di}{dt} = g(s,i) = 0 \quad (4)$$

Through Equation (4), the following fixed points are obtained for the SIR model:

$$P_1(s_{f1}, i_{f1}) = (0,1)$$
$$P_2(s_{f2}, i_{f2}) = \left(\frac{\gamma + \mu}{\beta}, \frac{\mu}{\gamma + \mu} - \frac{\mu}{\beta}\right) \quad (5)$$

The analysis of the dynamics of the SIR model given by [2], allows two states of equilibrium ($P_1$ and $P_2$) to determine. The fixed point ($P_1$), the population is free of infection and the fixed point $P_2$, the population of infected goes to an endemic balance. Endemic diseases are the result of long-term equilibrium between agent and host. There is a concern to adapt the parameters of the SIR model, so that the prediction of the evolution of the epidemic and simulation of the spread of the disease are satisfactory, due to the many uncertainties in the problem.

Applying the forward Euler scheme to model (3) can be obtained the following discrete-time SIR epidemic model

$$S_{n+1} = S_n + h(\mu - \mu S_n - \beta S_n I_n),$$
$$I_{n+1} = I_n + h(\beta I_n S_n - \gamma I_n - \mu I_n). \quad (6)$$

where $h$ is the step size, $\mu$, $\beta$ and $\gamma$ are defined as model (2).

**Individual-Based Model (IBM):** IBM's are important for both theory and application, because they allow researchers to consider aspects generally overlooked in analytical models such as, interaction between individuals, local interactions, complete life cycles, and so on. However, they have their weaknesses, the IBMs have their structure more complex than compartmental models, so they are more difficult to implement, analyze and understand. For this reason, Grimm and colleagues have developed a standard protocol for describing IBM. The protocol is composed of three blocks: overview, design concepts and details. The Overview block provides the general purpose and structure of the model. The Design concept block describes the general concepts underlying the design of the model. Finally, the Details block displays information that was omitted from the overview, such as initialization, input, and sub models. Background information in the ODD protocol can be obtained from [19,24]. An IBM allows to emulate simulation experiences in a computational environment taking into account the characteristics and interaction of each individual. In this way, aspects that are generally ignored by other models can be considered making the system more realistic. The purpose of IBM is to model infectious diseases, for this work is considered an equivalence between IBM and SIR, therefore IBM complements the SIR model, considering also those individuals can be divided into epidemiological states.

In the mathematical formulation considered for the IBM, the individual can be represented by

$$I_{n,t} = \begin{bmatrix} C_{n,1,t} & C_{n,2,t} & \ldots & C_{n,m,t} \end{bmatrix}, \quad (7)$$

where $n$ is a sequential number that identifies an individual, $m$ is the number of features, $t$ is the instant that the individual presents a specific set of characteristics and $C_{n,m,t}$ is a characteristic considered for the individual. The first characteristic is its epidemiological status, which may be susceptible, infected, recovered. Other characteristics may be age, duration of infection, sex, spatial location or any other characteristics of the individual considered relevant. It is important to highlight that the characteristics $C_{n,m,t}$ in (7) may change with time. A collection of individuals can be represented as a population:

$$P_t = \begin{bmatrix} I_{1,t} & I_{2,t} & I_{3,t} & \ldots & I_{m,t} \end{bmatrix}^T, \quad (8)$$

where $I_{m,t}$ is an individual at time $t$ and $P \in \mathbb{R}^{m \times n}$ is a matrix. Probabilistic distributions are used for an individual to verify when a state transition occurs. After the $I_{m,t}$ individuals are evaluated, the simulation time is incremented by $\Delta t$. This formulation of the model is quite generic, allowing to incorporate several characteristics of the individuals. The employed model has the following characteristics:

- $C_1 \in [0,1,2]$. That is, the individual may be in the susceptible, infected and retrieved state respectively.
- $C_2$ is the age of an individual, which receives an addition of $\Delta t$ in each transition time.
- $C_3$ is the maximum age at which the individual will live. At the moment of the birth of the individual this value is obtained by:

$$C_3 = -\mu \log(a_u) \quad (9)$$

where $\mu$ is the life expectancy of the population and $a_u$ is a random variance with uniform distribution, contained between 0 and 1.

- $C_4$ is the time that the individual is in the infecting state.
- $C_5$ is the maximum time the individual is in the infecting state. Like the $C_3$ characteristic, the maximum time the individual becomes infected is obtained a priori by:

$$C_5 = -\gamma \log(a_u) \quad (10)$$

where $\gamma$ is the infecting period.

Susceptible and recovered individuals do not present characteristics $C_4$ and $C_5$. In this case they are zero. The number of susceptible, infected and recovered individuals at each instant t of the population is denoted by $S_t$, $I_t$ and $R_t$, respectively.

Figure 2 shows an IBM flowchart. The characteristics of the initial population are determined randomly, given the probability distributions of the state variables.

## Results

This section presents the results of simulations experiments of IBM and SIR model.

**Developing an IBM for infectious diseases**: In order to show how IBM works and how it can be used as an alternative to the SIR model let us consider a hypothetical disease. Figure 3 shows IBM used to simulate a hypothetical disease that exhibits three epidemiological states: susceptible, infected and recovered. The parameters that have been used are $\Delta t = 0.1$, $\gamma = 1/3$, $\mu = 1/60$, $\alpha = 1/60$ and $\beta = 0.25$. The initial conditions are N (0) = 1000, S (0) = 900, I (0) = 10, and R (0) = 90. Figure 3 presents the results obtained from the IBM simulation. One can see that infected individuals tend to zero, which means that the epidemic tends to be eradicated. For this set of dynamic parameters IBM was simulated only once (Figure 3).

The same dynamic parameters have been considered to simulate the discretized SIR model. Figure 4 shows this result. It is possible to observe that the same behavior that have been seen in Figure 3 for IBM, occurs for this model, noting that infected individuals tend to zero, the epidemic tends to end (Figure 4).

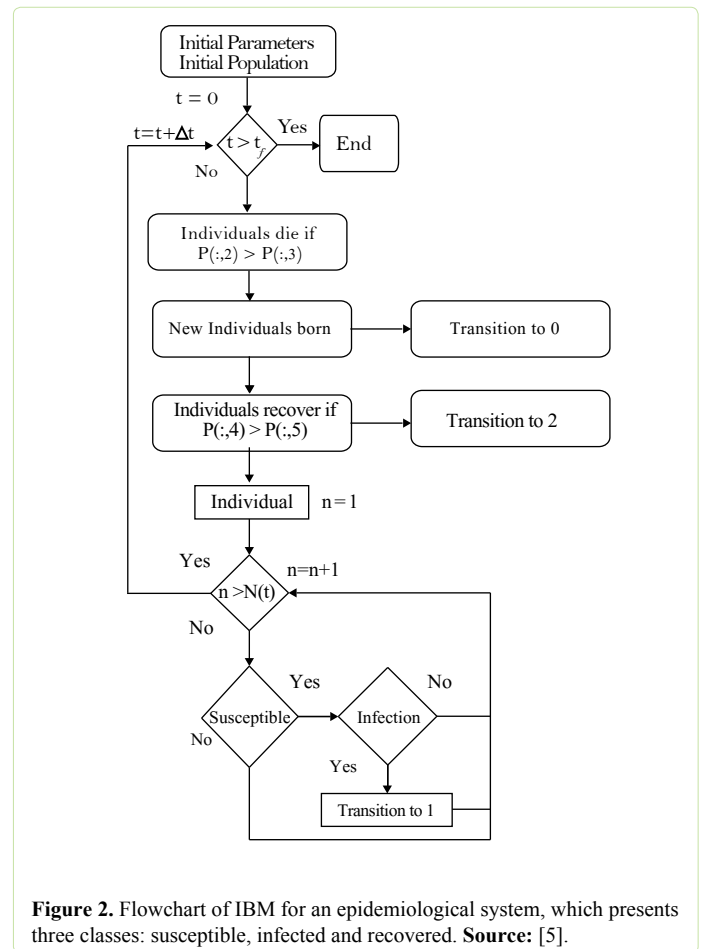

**Figure 2.** Flowchart of IBM for an epidemiological system, which presents three classes: susceptible, infected and recovered. **Source:** [5].

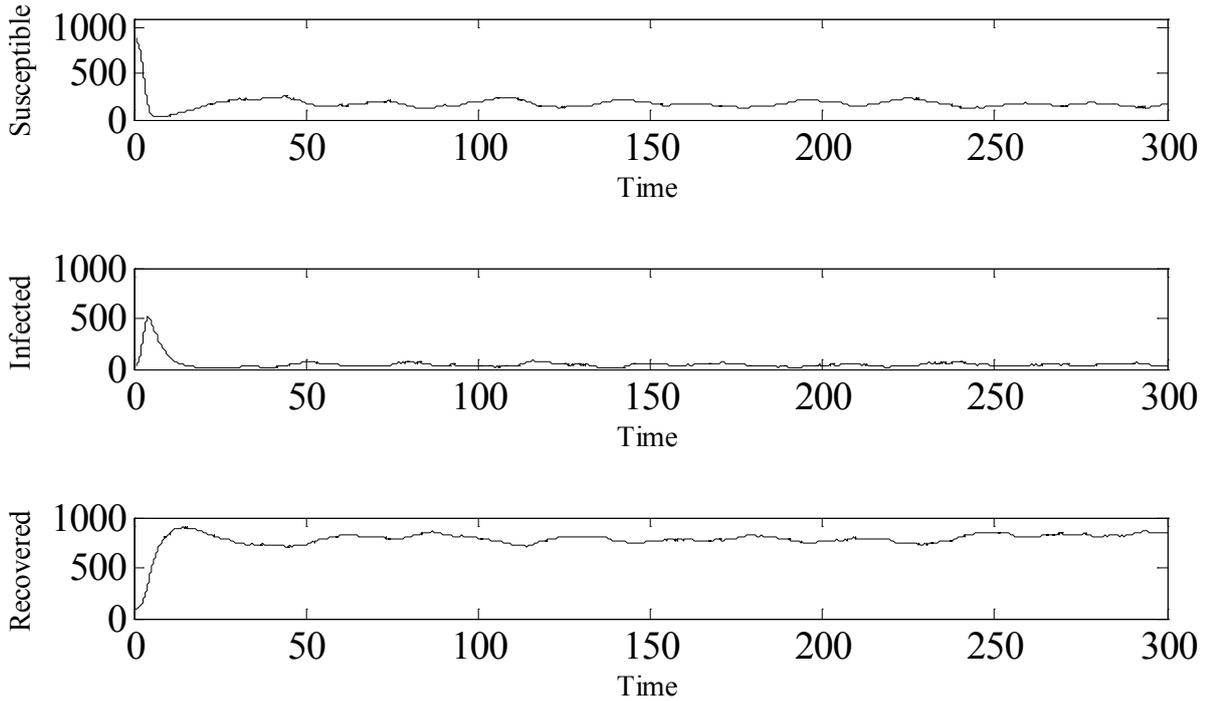

**Figure 3.** Simulation of IBM, showing Susceptible, Infected and Recovered individuals. $\Delta t = 0.1$, $\gamma = 1/3$, $\mu = 1/60$, $\alpha = 1/60$ and $\beta = 0.25$. The initial conditions are N (0) = 1000, S (0) = 900, I (0) = 10, and R (0) = 90.

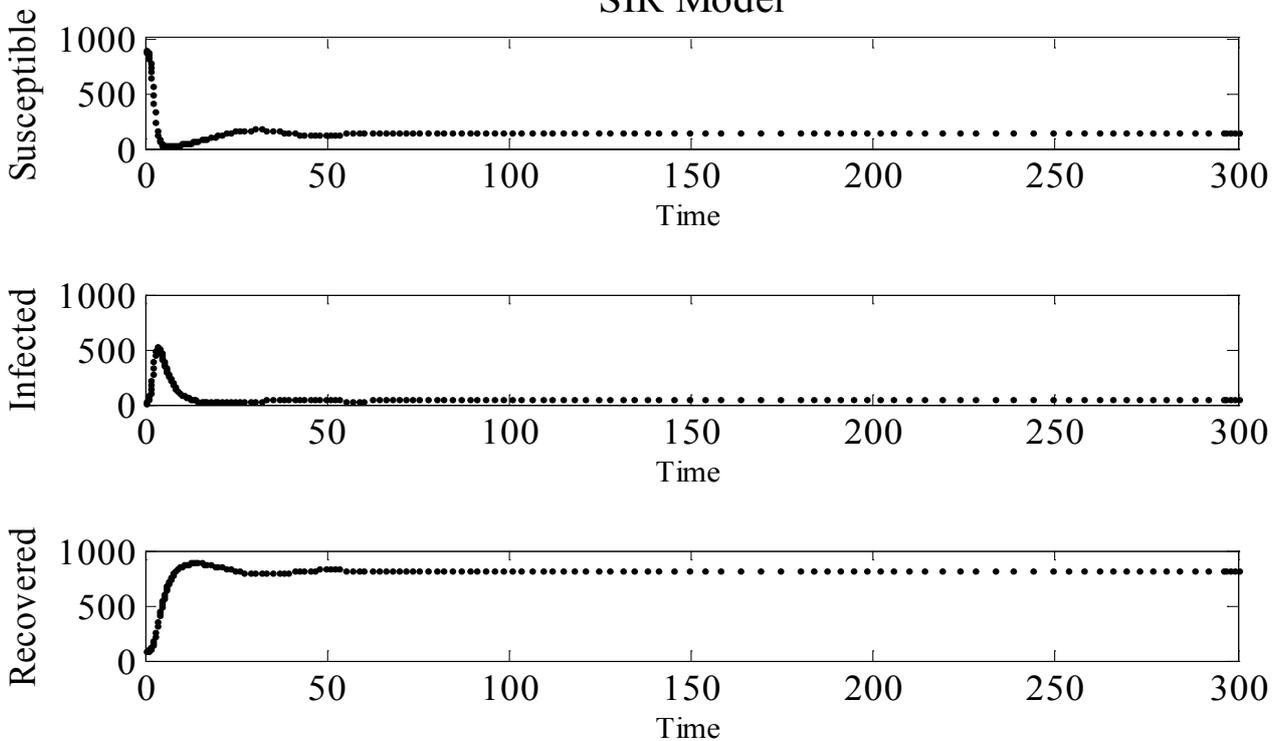

**Figure 4:** Simulation of SIR model, showing Susceptible, Infected and Recovered individuals. $\Delta t = 0.1$, $\gamma = 1/3$, $\mu = 1/60$, $\alpha = 1/60$ and $\beta = 0.25$. The initial conditions are N (0) = 1000, S (0) = 900, I (0) = 10, and R (0) = 90.

Next step is check for equivalence between the SIR model and IBM. To consolidate this result the Monte Carlo method have been used, where the IBM was simulated 100 times. Figure 5 depicts the Monte Carlo simulation for IBM, we notice a jump at $t = 230$, this is due to the stochasticity of the model (Figure 5).

To validate the equivalence between the SIR and IBM models the mean and standard deviation were calculated for each time value, as shown in Figure 6. It is possible to notice that IBM presents an average behavior that approaches the SIR model, represented in red. In this plot, it is presented only the number of infected individuals. Similar results are obtained for susceptible and recovered individuals.

It is important to show that by changing the integration step of the IBM model it loses stability, Figures 7-9 shows the simulated IBM model with the following parameters $\gamma = 1/3$, $\mu = 1/60$, $\alpha = 1/60$ and $\beta = 0.25$. The initial conditions are, N (0) = 1000, S (0) = 900, I (0) = 10, and R (0) = 90. The integration steps used for the simulations were $\Delta t = 0.1$, $\Delta t = 0.5$ and $\Delta t = 1$ (Figure 6-9).

For integration steps of $\Delta t = 0.5$ and $\Delta t = 1$, the IBM loses stability. The dynamic behavior of the model changes. The same behavior can be observed for compartmental models, for this case the discrete SIR model is studied, because its dynamic behavior is very plentiful and more complex than model SIR in continuous time [7]. That said, Figure 10 shows the bifurcation diagram for the SIR model. The following parameters for the discrete SIR were used $\mu = 0,2$ $\gamma = 0.15$ and $\beta = 0.12$ The initial conditions were $S(0) = 8$ and $I(0) = 5$. These parameters were taken from [7]. As $h$ varies a Flip bifurcation occurs, so it is easy to notice that there is chaos for this range of values (Figure 10).

## Conclusion

This article presented a literature review along with simulations of the SIR and IBM models for epidemiological systems. In the literature review, the selected papers were mostly on applications of diseases in humans. A total of 86 papers were selected, most of which addressed IBM applications on infectious diseases in humans. Some papers provided solutions for modeling and application of methods to prevent and control diseases using IBM as a computational tool. Examples using IBM have been shown in this paper, where the approach used of IBM allowed to consider specific characteristics of the individuals, considering them heterogeneous. The focus of this work was the analysis of phenomena associated with the models, both for continuous and discretized compartmental models and for stochastic models such as IBM. For IBM, some simulation experiments were tested, where the epidemiological system presents three classes: susceptible, infected and recovered. The population is considered constant and, at some point, integration steps are changed. The results show that IBM can model an epidemic leading to eradication, however, the model shows a loss of stability when the integration step is changed. A similar phenomenon occurs for the discrete SIR model, where for a specific set of values the model presents chaotic characteristics, which can be observed through the Bifurcation diagram. To consolidate these results it was necessary to prove the equivalence between the models, where IBM has an average behavior that approaches the SIR model. In addition, the number of infected individuals reaches an endemic value, which is expected for this simulation. Therefore, is important to note that, like deterministic models, stochastic models also have physical and structural limitations, i.e., chaos and loss of stability, leaving researchers alert to the use of these models even if they present reasonable results. Future directions of the IBM research on infectious diseases should be focused on the development of techniques to combat epidemics. In terms of application, the results presented here suggest that simulations using mathematical models, especially IBM, may be useful for evaluating methodologies for the eradication of several types of epidemics, however their limitations must be considered.

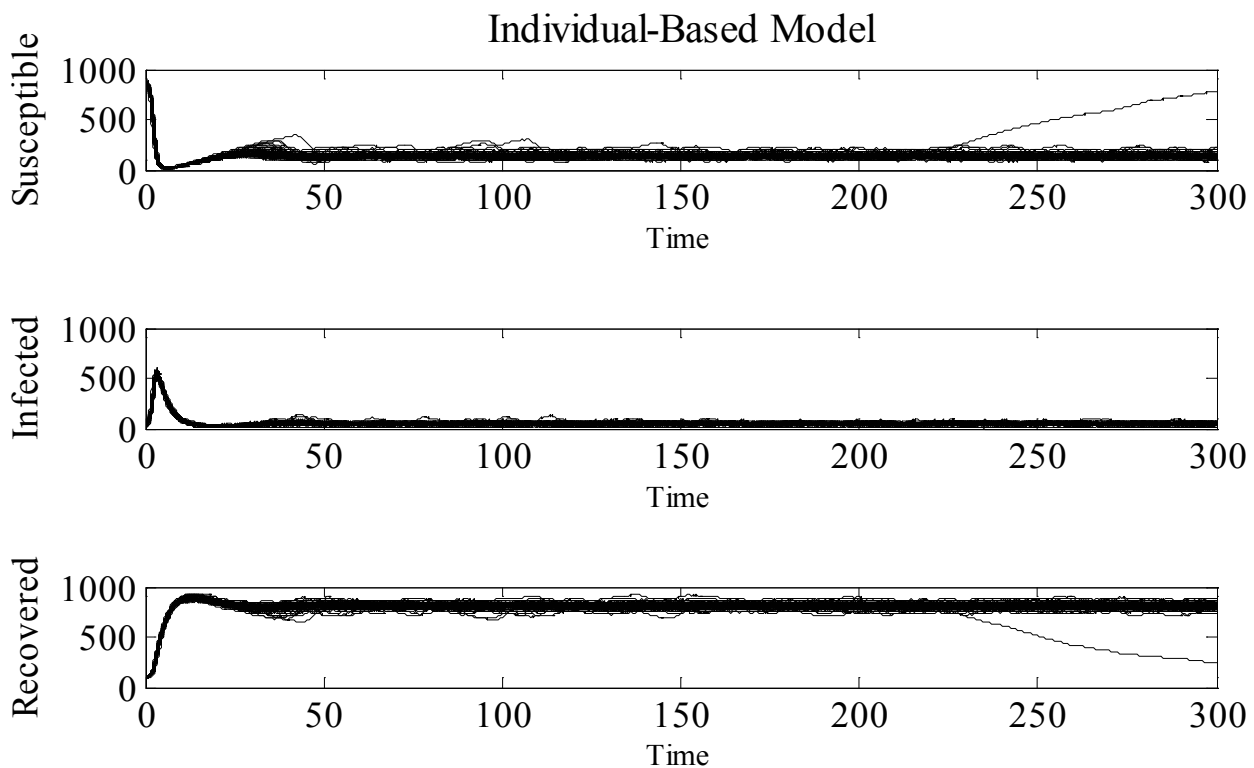

**Figure 5:** Monte Carlo simulation (100 runs). The parameters used $\Delta t = 0.1$, $\gamma = 1/3$, $\mu = 1/60$, $\alpha = 1/60$ and $\beta = 0.25$. The initial conditions are N (0) = 1000, S (0) = 900, I (0) = 10, and R (0) = 90. and R (0) = 90.

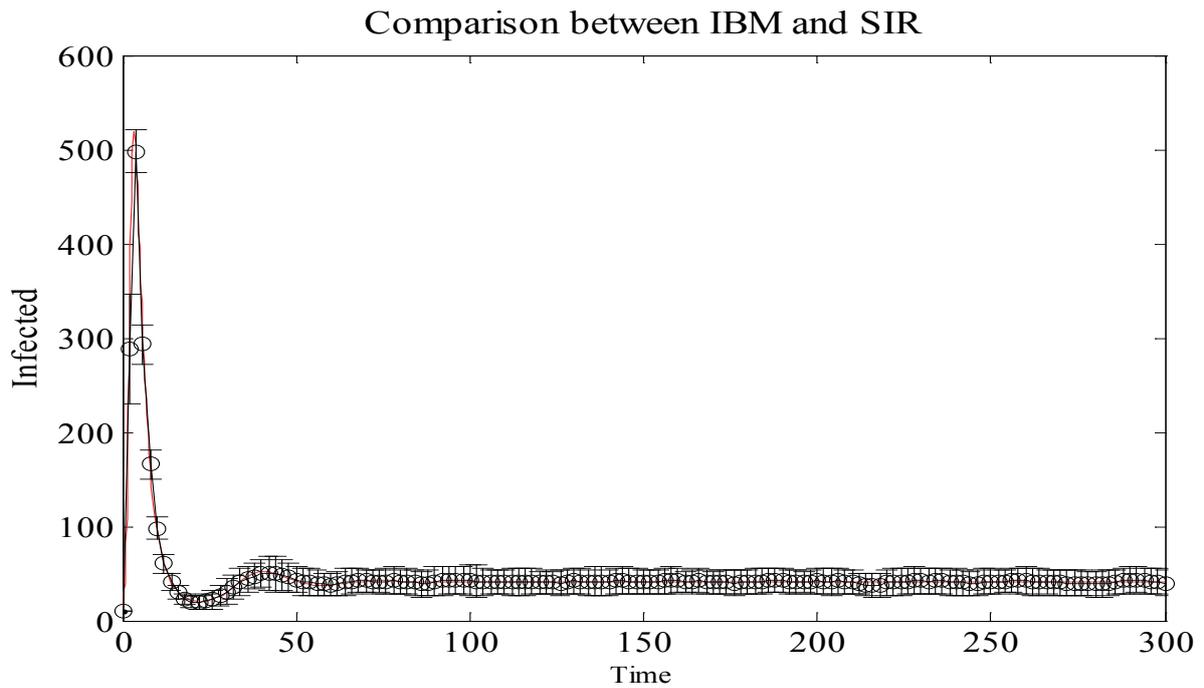

**Figure 6:** Monte Carlo simulation of IBM, showing average and standard deviation of Infected individual. In red individuals infected with the SIR model. The parameters used Δ$t$ = 0.1, $γ$ = 1/3, $μ$ = 1/60, $α$ = 1/60 and $β$ = 0.25. The initial conditions are N (0) = 1000, S (0) = 900, I (0) = 10, and R (0) = 90.

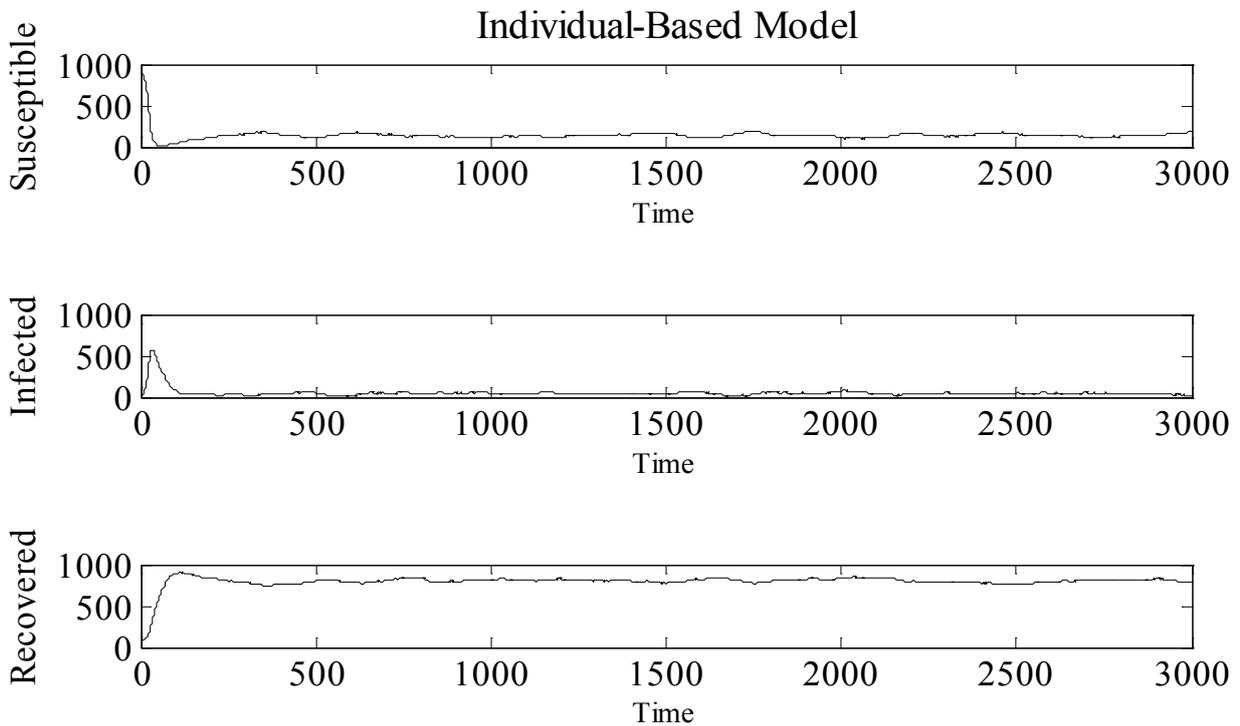

**Figure 7:** Simulation of IBM with integration step Δ$t$ = 0.1.

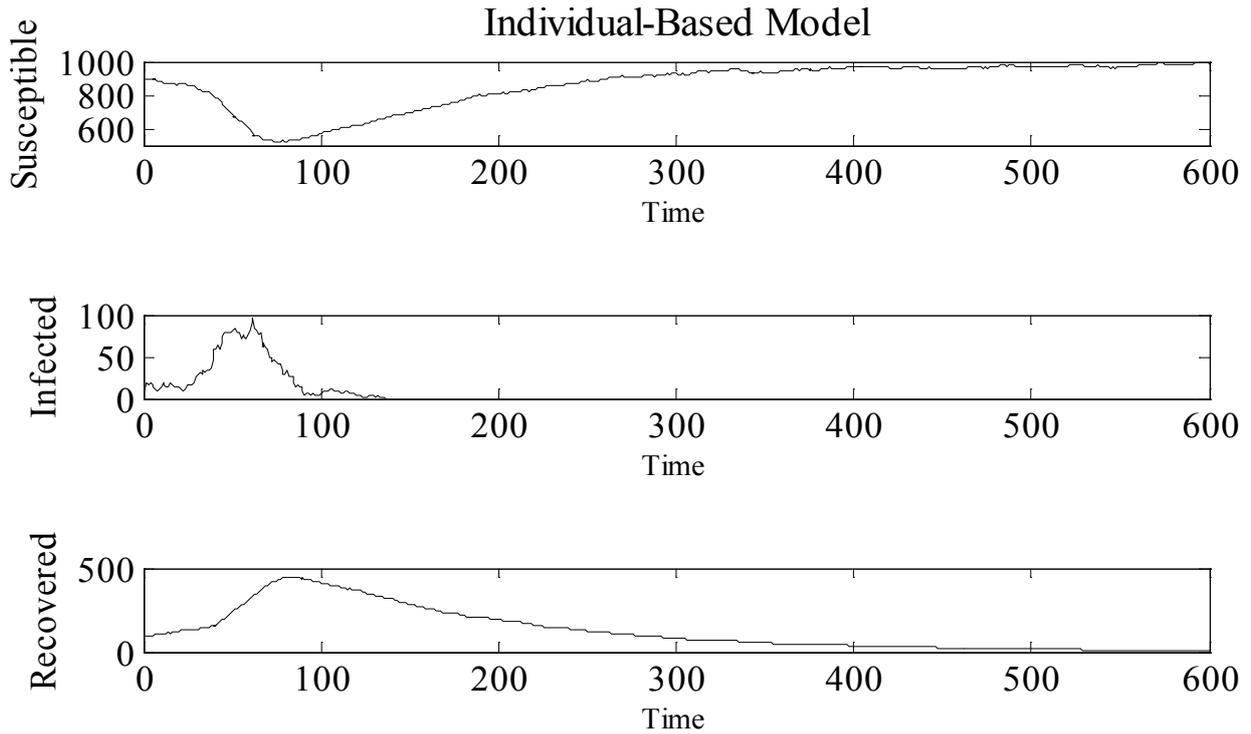

**Figure 8:** Simulation of IBM with integration step Δ*t* = 0.5.

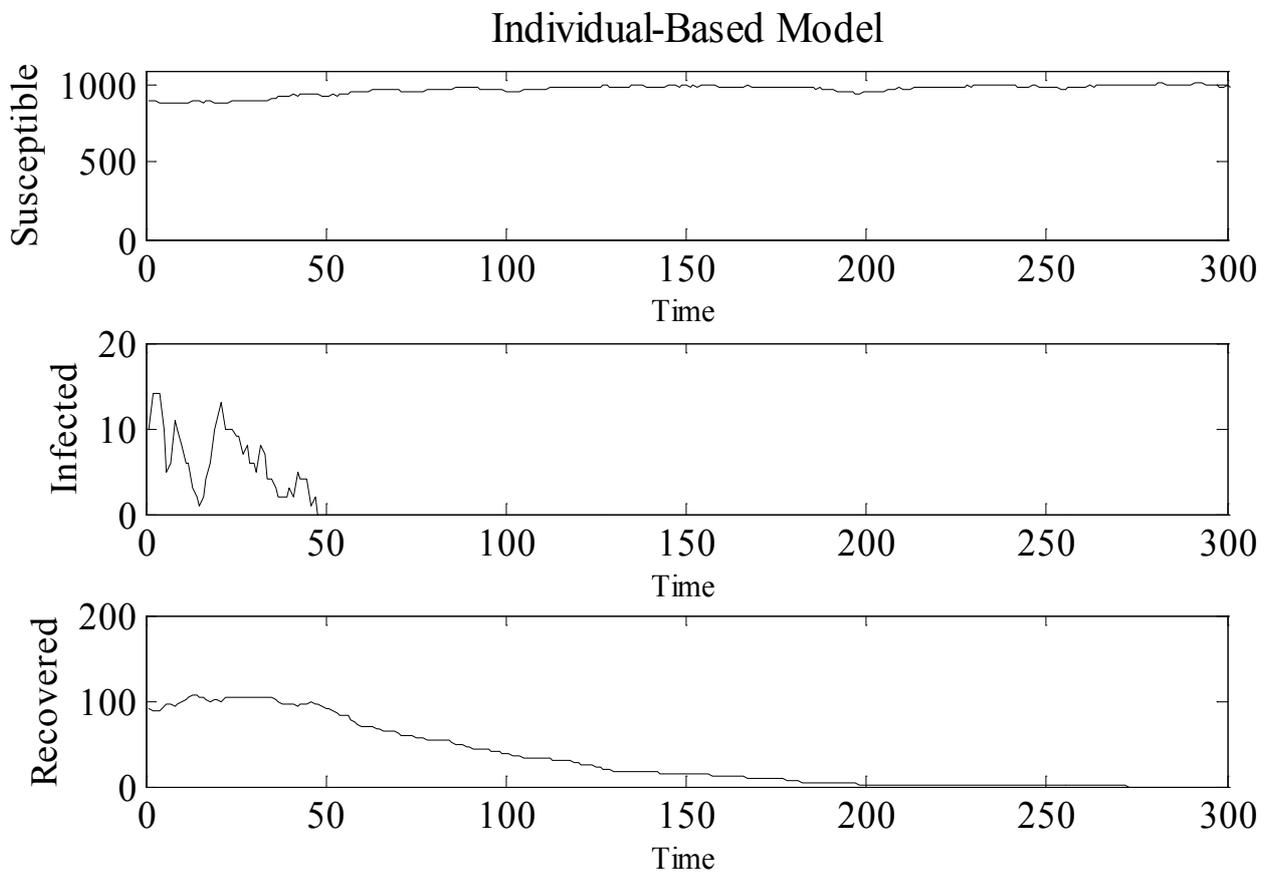

**Figure 9:** Simulation of IBM with integration step Δ*t* = 0.1.

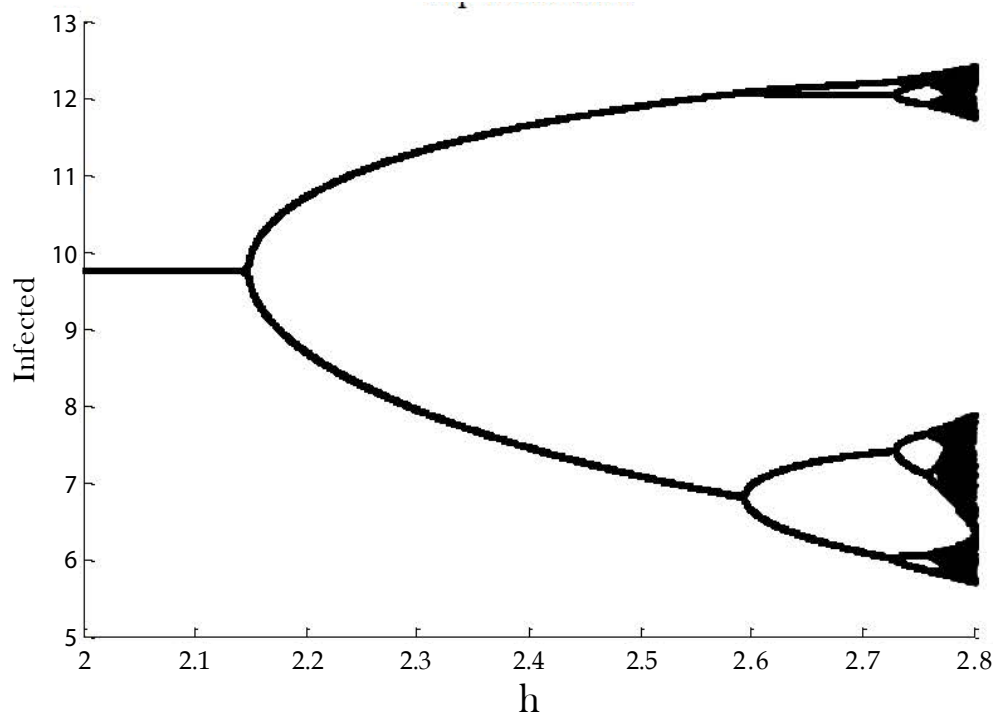

**Figure 10:** Discrete SIR model. The Flip bifurcation of I (n) with A = μ × N = 4, with μ = 0.2, γ = 0.15 and β = 0.12. h ϵ [2, 2.85] and initial values (S0, I0) = (8, 5). **Source:** [7].

## Acknowledgment

The authors thank CAPES, CNPq/INERGE and Fapemig for financial support.

## References

1. Morens DM, Folkers GK, Fauci AS (2004) The challenge of emerging and re-emerging infectious diseases. Nature 430: 242-249.

2. Anderson RM, May RM (1992) Infectious Diseases of Humans: Dynamics and Control. Oxford University Press, Oxford.

3. Kohn GC (1998) The Wordsworth Encyclopedia of Plague & Pestilence. Wordsworth Reference.

4. Pustil R (2016) Global AIDS. Aids 17: S3-11.

5. Nepomuceno EG, Takahashi RHC, Aguirre LA (2016) Individual Based-Model (IBM): An Alternative framework for epidemiological compartment models. Revista Brasileira de Biometria 34: 133-162.

6. Nepomuceno EG, Aguirre LA, Takahashi RHC (2018) Reducing vaccination level to eradicate a disease by means of a mixed control with isolation. Biomed Signal Process Control 40: 83-90.

7. Hu Z, Teng Z, Zhang L (2014) Stability and bifurcation analysis in a discrete SIR epidemic model. Math Comput Simul 97: 80-93.

8. Yang HM (2001) Epidemiologia matemática: estudo dos efeitos da vacinação em doenças de transmissão direta. Editora da Unicamp.

9. Kermack W, McKendrick A (1927) A contribution to the mathematical theory of epidemics. Proceedings of the Royal Society of London Series A: mathematical and physical Sciences 115: 700-721.

10. Allen LJS (1994) Some discrete-time SI, SIR, and SIS epidemic models. Math Biosci 124: 83-105.

11. Bjornstad ON, Finkenstadt BF, Grenfell BT (2002) Dynamics of measles epidemics: Estimating scaling of transmission rates using a Time series SIR model. Ecol Monogr 72: 169-184.

12. Stone L, Olinky R, Huppert A (2007) Seasonal dynamics of recurrent epidemics. Nature 446: 533-536.

13. Aron JL, Schwartz IB (1984) Seasonality and period-doubling bifurcations in an epidemic model. J Theor Biol 110: 665-679.

14. Nowak RM, Martin, May (1993) AIDS pathogenesis: mathematical models of HIV and SIV infections. Aids 7: S3-S18.

15. Black AJ, McKane AJ (2012) Stochastic formulation of ecological models and their applications. Trends Ecol Evol 27: 337-345.

16. Grimm V (1999) Ten years of individual-based modelling in ecology: What have we learned and what could we learn in the future?. Ecol Modell 115: 129-148.

17. Grimm V, Railsback S (2005) Individual-based Modeling and Ecology. Princeton Series in Theoretical and Computational Biology.

18. Omer SB, Salmon DA, Orenstein WA, DeHart MP, Halsey N (2009) Vaccine Refusal, Mandatory Immunization, and the Risks of Vaccine-Preventable Diseases. N Engl J Med 360: 981-1988.

19. Grimm V, Berger U, DeAngelis DL, Polhill JG, Giske J, et al. (2010) The ODD protocol: A review and first update. Ecol Modell 221: 2760-2768.

20. Tisue U, Seth, Wilensky (2004) NetLogo: Design and implementation of a multi-agent modeling environment. International conference on complex systems 21: 16-21.

21. Railsback V, Steven F and Grimm (2011) Agent-based and individual-based modeling: a practical introduction. Princeton University Press.

22. DeAngelis DL, Grimm V (2014) Individual-based models in ecology after four decades. F1000Prime Reports.

23. Willem L, Verelst F, Bilcke J, Hens N, Beutels P (2017) Lessons from a decade of individual-based models for infectious disease transmission: a systematic review (2006-2015). BMC Infect Dis 17: 612.

24. Grimm V (2006) A standard protocol for describing individual-based and agent-based models. Ecol Modell 198: 115-126.

25. Yan S (2015) The model of Ebola disease control and optimization algorithm of pharmaceutical logistics path. 8th International Conference on Biomedical Engineering and Informatics (BMEI).

26. Pell B, Kuang Y, Viboud C, Chowell G (2016) Using phenomenological models for forecasting the 2015 Ebola challenge. Epidemics.

27. Kurahashi S, Terano T (2015) A Health Policy Simulation Model of Smallpox and Ebola Haemorrhagic Fever. Agent and Multi-Agent Systems: Technologies and Applications. Springer.

28. Bissett K, Cadena J, Khan M, Kuhlman CJ, Lewis B, et al. (2016) An integrated agent-based approach for modeling disease spread in large populations to support health informatics. International Conference on Biomedical and Health Informatics (BHI).

29. Getz WM, Salter RM, Sippl-Swezey N (2015) Using Nova to construct agent-based models for epidemiological teaching and research. Winter Simulation Conference (WSC) IEEE.

30. Sanchez PJ, Sanchez SM (2015) A scalable discrete event stochastic agent-based model of infectious disease propagation. Winter Simulation Conference (WSC), IEEE.

31. Ciavarella C, Fumanelli L, Merler S, Cattuto C, Ajelli M (2016) School closure policies at municipality level for mitigating influenza spread: a model-based evaluation. BMC Infect Dis 16: 576.

32. Halder N, Kelso JK, Milne GJ (2010) Analysis of the effectiveness of interventions used during the 2009 A/H1N1 influenza pandemic. BMC Public Health 10: 168.

33. Smieszek T, Balmer M, Hattendorf J, Axhausen KW, Zinsstag J et al. (2011) Reconstructing the 2003/2004 H3N2 influenza epidemic in Switzer-land with a spatially explicit, individual-based model. BMC Infect Dis 11: 115.

34. Milne GJ, Kelso JK, Kelly HA, Huband ST, McVernon J (2008) A Small Community Model for the Transmission of Infectious Diseases: Comparison of School Closure as an Intervention in Individual-Based Models of an Influenza Pandemic. PLoS ONE 3: e4005.

35. CiofidegliAtti ML, Merler S, Rizzo C, Ajelli M, Massari M, et al. (2008) Mitigation Measures for Pandemic Influenza in Italy: An Individual Based Model Considering Different Scenarios. PLoS ONE 3: e1790.

36. Morimoto T, Ishikawa H (2010) Assessment of intervention strategies against a novel influenza epidemic using an individual-based model. Environ Health Prev Med 15: 151-161.

37. Roche B, Drake JM, Rohani P (2011) An Agent-Based Model to study the epidemiological and evolutionary dynamics of Influenza viruses. BMC Bioinforma 12: 87.

38. Hyder A, Buckeridge DL, Leung B (2013) Predictive Validation of an Influenza Spread Model. PLoS ONE 8: e65459.

39. Sukumar SR, Nutaro JJ (2012) Agent-Based vs. Equation-Based Epidemiological Models: A Model Selection Case Study. International Conference on BioMedical Computing (BioMedCom), IEEE.

40. Liu Y (2010) Investigation of Prediction and Establishment of SIR Model for H1N1 Epidemic Disease. 4th International Conference on Bioinformatics and Biomedical Engineering, IEEE.

41. Amouroux E, Gaudou B, Desvaux S, Drogoul A (2010) O.D.D.: A Promising but Incomplete Formalism for Individual-Based Model Specification. International Conference on Computing & Communication Technologies, Research, Innovation, and Vision for the Future (RIVF), IEEE.

42. Ponnambalam L, Xiuju F, Goh RSM, Sarawgi D (2016) A multi-agent model for adaptive vaccination during infectious disease outbreaks. International Conference on Computing Technologies and Intelligent Data Engineering (ICC-TIDE'16), IEEE.

43. Ge, Yuanzheng, et al. (2018) A synthetic computational environment: To control the spread of respiratory infections in a virtual university. Physica A: Statistical Mechanics and its Applications 492: 93-104.

44. Guizani N, Ghafoor A (2014) Modeling and evaluation of disease spread behaviors. International Wireless Communications and Mobile Computing Conference (IWCMC), IEEE.

45. da Cruz AR, Cardoso RT, Takahashi RH (2017) Multiobjective synthesis of robust vaccination policies. Appl Soft Comput 50: 34-47.

46. Duan W, Qiu X, Cao Z, Zheng X, Cui K (2013) Heterogeneous and Stochastic Agent-Based Models for Analyzing Infectious Diseases' Super Spreaders. IEEE Intelligent Systems 28: 18-25.

47. Jha SK, Ramanathan A (2012) Quantifying Uncertainty in Epidemiological Models. International Conference on BioMedical Computing (BioMedCom), IEEE.

48. Saravanan M, Karthikeyan P, Arathi A, Kiruthika M, Suganya S (2013) Mobile agent-based approach for modeling the epidemics of communicable diseases. International Conference on Advances in Social Networks Analysis and Mining, New York, USA.

49. Naron S, Wasserkrug S (2007) Utilizing model characteristics to obtain efficient parallelization in the context of agent based epidemiological models. Winter Simulation Conference, IEEE.

50. Ramanathan A, Steed CA, Pullum LL (2012) Verification of Compartmental Epidemiological Models Using Metamorphic Testing, Model Checking and Visual Analytics. International Conference on BioMedical Computing (BioMedCom), IEEE.

51. Qin H, Shapiro A, Yang L (2012) Emerging Infectious Disease: A Computational Multi-agent Model. International Conference on BioMedical Computing (BioMedCom), IEEE.

52. Piau P, Siang AK, Labadin J (2013) An epidemic aggregate and individual-based model in a patchy environment. 8th International Conference on Information Technology in Asia (CITA), IEEE.

53. Kuylen E, Stijven S, Broeckhove J, Willem L (2017) Social Contact Patterns in an Individual-based Simulator for the Transmission of Infectious Diseases (Stride). Procedia Comput Sci 108: 2438-2442.

54. Monteiro JFG, Escudero DJ, Weinreb C, Flanigan T, Galea S, et al. (2016) Understanding the effects of different HIV transmission models in individual-based microsimulation of HIV epidemic dynamics in people who inject drugs. Epidemiol Infect 144: 1683-1700.

55. Monteiro JFG, Galea S, Flanigan T, Monteiro MdL, Friedman SR, et al. (2015) Evaluating HIV prevention strategies for populations in key affected groups: the example of Cabo Verde. Int J Public Health 60: 457-466.

56. Smith JA, Heffron R, Butler AR, Celum C, Baeten JM et al. Could misreporting of condom use explain the observed association between injectable hormonal contraceptives and HIV acquisition risk?. Contraception 95: 424-430.

57. Demongeot J, Hansen O, Jannot A, Taramasco C (2012) Random Modelling of Contagious (Social and Infectious) Diseases: Examples of Obesity and HIV and Perspectives Using Social Networks. 26th International Conference on Advanced Information Networking and Applications Workshops, IEEE.

58. Alam SJ, Zhang X, Romero-Severson EO, Henry C, Zhong L, et al. (2013) Detectable signals of episodic risk effects on acute HIV transmission: Strategies for analyzing transmission systems using genetic data. Epidemics 5: 44-55.

59. Gray RT, Prestage GP, Down I, Ghaus MH, Hoare A, et al. (2013) Increased HIV Testing Will Modestly Reduce HIV Incidence among Gay Men in NSW and Would Be Acceptable if HIV Testing Becomes Convenient. PLoS ONE 8: e55449.

60. de Vos AS, Prins M, Coutinho RA, van der Helm JJ, et al. (2014) Treatment as prevention among injecting drug users; extrapolating from the Amsterdam cohort study. AIDS 28: 911-918.

61. Kim JH, Riolo RL, Koopman JS (2010) HIV Transmission by Stage of Infection and Pattern of Sexual Partnerships. Epidemiology 21: 676-684.

62. Dimitrov D, Donnell D, Brown ER (2015) High Incidence Is Not High Exposure: What Proportion of Prevention Trial Participants Are Exposed to HIV?. PLOS ONE 10: e0115528.

63. Smith JA, Sharma M, Levin C, Baeten JM, van Rooyen H, et al. (2015) Cost-effectiveness of community-based strategies to strengthen the continuum of HIV care in rural South Africa: a health economic modelling analysis. The Lancet HIV 2: e159–e168.

64. White PJ, Fox J, Weber J, Fidler S, Ward H (2014) How Many HIV Infections May Be Averted by Targeting Primary Infection in Men Who Have Sex With Men? Quantification of Changes in Transmission-Risk Behavior, Using an Individual-Based Model. Int J Infect Dis 210: S594-S599.

65. Leonenko VN, Pertsev NV, Artzrouni M (2015) Using High Performance Algorithms for the Hybrid Simulation of Disease Dynamics on CPU and GPU. Procedia Comput Sci 51:150-159.

66. Andrianakis I, Vernon I, McCreesh N, McKinley TJ, Oakley JE, et al. (2017) History matching of a complex epidemiological model of human immunodeficiency virus transmission by using variance emulation. J R Stat Soc Ser C Appl Stat 66: 717-740.

67. Kasaie P, Pennington J, Shah MS, Berry SA, German D, et al. (2017) The Impact of Preexposure Prophylaxis Among Men Who Have Sex With Men. J Acquir Immune Defic Syndr 75: 175-183.

68. van Santen DK, de Vos AS, Matser A, Willemse SB, Lindenburg K, et al. (2016) Cost-Effectiveness of Hepatitis C Treatment for People Who Inject Drugs and the Impact of the Type of Epidemic; Extrapolating from Amsterdam, the Netherlands. PLOS ONE 11: e0163488.

69. Althaus CL, Turner KME, Schmid BV, Heijne JCM, Kretzschmar M, et al. (2012) Transmission of Chlamydia trachomatis through sexual partnerships: a comparison between three individual-based models and empirical data. J R Soc Interface 9: 136-146.

70. Kasaie P, Mathema B, Kelton WD, Azman AS, Pennington J, et al. (2015) A Novel Tool Improves Existing Estimates of Recent Tuberculosis Transmission in Settings of Sparse Data Collection. PLOS ONE 10: e0144137.

71. Shrestha S, Hill AN, Marks SM, Dowdy DW (2017) Comparing Drivers and Dynamics of Tuberculosis in California, Florida, New York, and Texas. Am J Respir Crit Care Med 196: 1050-1059.

72. Guzzetta G, Ajelli M, Yang Z, Merler S, Furlanello C, et al. (2011) Modeling socio-demography to capture tuberculosis transmission dynamics in a low burden setting. J Theor Biol 289: 197-205.

73. Denholm JT, McBryde ES (2014) Can Australia eliminate TB? Modelling immigration strategies for reaching MDG targets in a low-transmission setting. Aust New Zealand J Public Health 38: 78-82.

74. Prats C, Montañola-Sales C, Gilabert-Navarro JF, Valls J, Casanovas-Garcia J, et al. (2016) Individual-Based Modeling of Tuberculosis in a User-Friendly Interface: Understanding the Epidemiological Role of Population Heterogeneity in a City. Frontiers in Microbiology.

75. Kasaie P, Dowdy DW, Kelton WD (2014) Estimating the proportion of tuberculosis recent transmission via simulation. Winter Simulation Conference 2014, IEEE.

76. Chao DL, Longini IM, Halloran ME (2013) The Effects of Vector Movement and Distribution in a Mathematical Model of Dengue Transmission. PLoS ONE 8: e76044.

77. Favier C, Schmit D, Muller-Graf CD, Cazelles B, Degallier N, et al. (2005) Influence of spatial heterogeneity on an emerging infectious disease: the case of dengue epidemics. Proc R Soc Lond B Biol Sci 272: 1171-1177.

78. de Lima T, Lana R, de Senna Carneiro T, Codeço C, Machado G, et al. (2016) DengueME: A Tool for the Modeling and Simulation of Dengue Spatiotemporal Dynamics. Int J Environ Res Public Health 13: 920.

79. Lima T, Carneiro T, Silva L, Lana R, Codeço C, et al. (2014) A framework for modeling and simulating Aedesaegypti and dengue fever dynamics. Winter Simulation Conference 2014, IEEE.

80. Otero M, Barmak DH, Dorso CO, Solari HG, Natiello MA (2011) Modeling dengue outbreaks. Math Biosci 232: 87-95.

81. Hahn JA, Wylie D, Dill J, Sanchez MS, Lloyd-Smith JO, et al. (2009) Potential impact of vaccination on the hepatitis C virus epidemic in injection drug users. Epidemics 1: 47-57.

82. Wong WW, Feng ZZ, Thein HH (2016) A parallel sliding region algorithm to make agent-based modeling possible for a large-scale simulation: Modeling Hepatitis C epidemics in Canada. IEEE J Biomed and Health Inform 20: 1538-1544.

83. Cousien A, Tran VC, Deuffic-Burban S, Jauffret-Roustide M, Dhersin JS, et al (2015) Dynamic modelling of hepatitis C virus transmission among people who inject drugs: a methodological review. J Viral Hepat 22: 213-229.

84. Rolls DA, Sacks-Davis R, Jenkinson R, McBryde E, Pattison P, et al. (2013) Hepatitis C Transmission and Treatment in Contact Networks of People Who Inject Drugs. PLoS ONE 8: e78286.

85. Cousien A, Tran VC, Deuffic-Burban S, Jauffret-Roustide M, Dhersin JS, et al (2016) Hepatitis C treatment as prevention of viral transmission and liver-related morbidity in persons who inject drugs. Hepatol 63: 1090-1101.

86. Gountas I, Sypsa V, Anagnostou O, Martin N, Vickerman P, et al. Treatment and primary prevention in people who inject drugs for chronic hepatitis C infection: is elimination possible in a high-prevalence setting?. Addiction 112: 1290-1299.

87. Schumaker, Nathan H, Allen B (2018) HexSim: a modeling environment for ecology and conservation. Landsc Ecol 33: 1-15.

88. Rebaudo F, Costa J, Almeida CE, Silvain JF, Harry M, et al. (2014) Simulating Population Genetics of Pathogen Vectors in Changing Landscapes: Guidelines and Application with Triatomabrasiliensis. PLoS Negl Trop Dis 8: e3068.

89. Spear RC, Wang S (2015) Exploring the Contribution of Host Susceptibility to Epidemiological Patterns of Schistosoma japonicum Infection Using an Individual-Based Model. Am J Trop Med Hyg 92: 1245-1252.

90. Laperriere V, Brugger K, Rubel F (2016) Cross-scale modeling of a vector-borne disease, from the individual to the metapopulation: The seasonal dynamics of sylvatic plague in Kazakhstan. Ecol Modell 342: 34-48.

91. Hui BB, Whiley DM, Donovan B, Law MG, Regan DG (2017) Identifying factors that lead to the persistence of imported gonorrhoeae strains: a modelling study. Sex Transm Infect 93: 221-225.

92. Campbell PT, McVernon J, Geard N (2017) Determining the Best Strategies for Maternally Targeted Pertussis Vaccination Using an Individual-Based Model. Am J Epidemiol 186: 109-117.

93. Blok DJ, De Vlas SJ, Richardus JH (2015) Global elimination of leprosy by 2020: are we on track?. Parasit Vectors 8: 548.

94. de Matos HJ, Blok DJ, de Vlas SJ, Richardus JH (2016) Leprosy New Case Detection Trends and the Future Effect of Preventive Interventions in Pará State, Brazil: A Modelling Study. PLoS Negl Trop Dis 10: e0004507.

95. Blok DJ, de Vlas SJ, Fischer EA, Richardus JH (2015) Mathematical Modelling of Leprosy and Its Control. Adv Parasitol 87: 33-51.

96. Depinay JMO, Mbogo CM, Killeen G, Knols B, Beier J, et al. (2004) A simulation model of African Anopheles ecology and population dynamics for the analysis of malaria transmission. Malar J 3: 29.

97. Gu W, Killeen GF, Mbogo CM, Regens JL, Githure JI, et al. (2003) An individual-based model of Plasmodium falciparum malaria transmission on the coast of Kenya. Trans R Soc Trop Med Hyg 97: 43-50.

98. Dommar CJ, Lowe R, Robinson M, Rodo X (2014) An agent-based model driven by tropical rainfall to understand the spatio-temporal heterogeneity of a chikungunya outbreak. Acta Trop 129: 61-73.

99. Silhol R, Boelle PY (2011) Modelling the Effects of Population Structure on Childhood Disease: The Case of Varicella. PLoS Comput Biol 7: e1002105.

100. Ogunjimi B, Willem L, Beutels P, Hens N (2015) Integrating between-host transmission and within-host immunity to analyze the impact of varicella vaccination on zoster. eLife 4: e07116.

101. Ajelli M, Merler S (2009) An individual-based model of hepatitis A transmission. J Theor Biol 259: 478-488.

102. Tian T (2011) Stochastic dynamics of a Hepatitis B virus transmission model. International Symposium on IT in Medicine and Education, IEEE.

103. Vanni T, Luz PM, Foss A, Mesa-Frias M, Legood R (2012) Economic modelling assessment of the HPV quadrivalent vaccine in Brazil: A dynamic individual-based approach. Vaccine 30: 4866-4871.

104. Choi YH, Chapman R, Gay N, Jit and M (2012) Potential overestimation of HPV vaccine impact due to unmasking of non-vaccine types: Quantification using a multi-type mathematical model. Vaccine 30: 3383-3388.

105. Ben HadjYahia MB, Jouin-Bortolotti A, Dervaux B (2015) Extending the Hu-man Papillomavirus Vaccination Programme to Include Males in High-Income Countries: A Systematic Review of the Cost-Effectiveness Studies. Clin Drug Investig 35: 471-485.

106. Thompson KM, Kisjes KH (2015) Modeling Measles Transmission in the North American Amish and Options for Outbreak Response. Risk Anal 36: 1404-1417.

107. Kim JH, Rho SH (2015) Transmission dynamics of oral polio vaccine viruses and vaccine-derived polioviruses on networks. J Theor Biol 364: 266-274.

108. Kisjes KH, DuintjerTebbens RJ, Wallace GS, Pallansch MA, Cochi SL, et al. (2014) Individual-Based Modeling of Potential Poliovirus Transmission in Connected Religious Communities in North America With Low Uptake of Vaccination. Int J Infect Dis 210: S424-S433.

109. Poletti P, Merler S, Ajelli M, Manfredi P, Munywoki PK, et al. (2015) Evaluating vaccination strategies for reducing infant respiratory syncytial virus infection in low-income settings. BMC Med 13: 49.

110. Zhou J, Gong J, Li (2006) Human Daily Behavior Based Simulation for Epidemic Transmission: A Case Study of SARS. 16th International Conference on Artificial Reality and Telexistence-Workshops (ICAT'06), IEEE.

111. Zenihana T, Ishikawa H (2010) Effectiveness assessment of countermeasures against bioterrorist smallpox attacks in Japan using an individual-based model. Environ Health Prev Med 15: 84-93.

112. Sato H, Sakurai Y (2012) The Contribution of Residents Who Cooperate With Ring-Vaccination Measures Against Smallpox Epidemic. Disaster Med Public Health Prep 6: 270-276.